\documentclass[10pt,a4paper,twocolumn,oneside]{article}
\usepackage[a4paper,tmargin=25mm,bmargin=25mm,lmargin=25mm,rmargin=25mm]{geometry}
\usepackage{fontenc}[T1]
\usepackage{alltt}
\usepackage{graphicx}

\usepackage[english]{babel}
\usepackage{url}
\usepackage{multirow}
\usepackage{eurosym} 
\graphicspath{{figures/}}
\usepackage{subfig}
\usepackage{hyperref}
\usepackage{amsthm}
	\newtheorem{mydef}{Definition}
\usepackage{kbordermatrix}
\usepackage[ruled,vlined,linesnumbered]{algorithm2e}
\usepackage{verbatim}
\usepackage{multirow}
\usepackage{booktabs}
\usepackage{amsmath}
\usepackage{hyperref}
\usepackage{breakurl}
\usepackage{balance}

\begin{document}






\title{Malware Classification based on Call Graph Clustering\footnote{This research has been supported by TEKES -- the Finnish Funding Agency for Technology and Innovation as part of its ICT SHOK Future Internet research programme, grant 40212/09.}}

\author{Joris Kinable\\
	Orestis Kostakis\\
	Aalto University, Department of Information and Computer Science \\
	Helsinki Institute for Information Technology\\
	P.O. Box 15400, FI-00076 Aalto, Finland\\
	\texttt{Joris.Kinable@tkk.fi, Orestis.Kostakis@tkk.fi}}
\date{\today}
\maketitle

\begin{abstract}
Each day, anti-virus companies receive tens of thousands samples of potentially harmful executables. Many of the malicious samples are variations of previously encountered malware, created by their authors to evade pattern-based detection. Dealing with these large amounts of data requires robust, automatic detection approaches.\\
This paper studies malware classification based on \emph{call graph clustering}. By representing malware samples as call graphs, it is possible to abstract certain variations away, and enable the detection of structural similarities between samples. The ability to cluster similar samples together will make more generic detection techniques possible, thereby targeting the commonalities of the samples within a cluster.\\
To compare call graphs mutually, we compute pairwise graph similarity scores via graph matchings which approximately minimize the \emph{graph edit distance}. Next, to facilitate the discovery of similar malware samples, we employ several clustering algorithms, including \emph{$k$-medoids} and \emph{DBSCAN}. Clustering experiments are conducted on a collection of real malware samples, and the results are evaluated against manual classifications provided by human malware analysts.\\
Experiments show that it is indeed possible to accurately detect malware families via call graph clustering. We anticipate that in the future, call graphs can be used to analyse the emergence of new malware families, and ultimately to automate implementation of generic detection schemes.

\vspace{3mm}
\end{abstract}


\section{Introduction}
\label{sec:Introduction}

Tens of thousands of potentially harmfull executables are submitted for analysis to data security companies on a daily basis. To deal with these vast amounts of samples in a timely manner, autonomous systems for detection, identification and categorization are required. However, in practice automated detection of malware is hindered by code obfuscation techniques such as packing or encryption of the executable code. Furthermore, cyber criminals constantly develop new versions of their malicious software to evade pattern-based detection by anti-virus products \cite{symThreadReport}.\\
For each sample a data security company receives, it has to be determined whether the sample is malicious or has been encountered before, possibly in a modified form. Analogous to the human immune system, the ability to recognize commonalities among malware which belong to the same malware family would allow anti-virus products to proactively detect both known samples, as well as future releases of the malware samples from the family. To facilitate the recognition of similar samples or commonalities among multiple samples which have been subject to change, a high-level structure, i.e.\ an abstraction, of the samples is required. One such abstraction is the \emph{call graph}. A call graph is a graphical representation of a binary executable in which functions are modeled as vertices, and calls between those functions as directed edges \cite{callGraph}.\\
This paper deals with mutual comparisons of malware via their call graph representations, and the classification of structurally similar samples into malware families through the use of clustering algorithms. So far, only a limited amount of research has been devoted to malware classification and identification using graph representations. Flake \cite{FLAK04} and later Dullien and Rolles \cite{DURO05} describe approaches to finding subgraph isomorphisms within control flow graphs, by mapping functions from one flow graph to the other. Functions which could not be reliably mapped have been subject to change. Via this approach, the authors of both papers can for instance reveal differences between versions of the same executable or detect code theft. Additionally, the authors of \cite{DURO05} suggest that security experts could save valuable time by only analyzing the differences among variants of the same malware.\\
Preliminary work on call graphs specifically in the context of malware analysis has been performed by Carrera and Erd\'{e}lyi \cite{CAER04}. To speed up the process of malware analysis, Carrera and Erd\'{e}lyi use call graphs to reveal similarities among multiple malware samples. Furthermore, after deriving similarity metrics to compare call graphs mutually, they apply the metrics to create a small malware taxonomy using a hierarchical clustering algorithm. Briones and Gomez \cite{BRGO08} continued the work started by Carrera and Erd\'{e}lyi. Their contributions mainly focus on the design of a distributed system to compare, analyse and store call graphs for automated malware classification. Finally the first large scale experiments on malware comparisons using real malware samples were recently published in \cite{hcs09,KIN10}. Additionally, the authors of \cite{hcs09} describe techniques for efficient indexing of call graphs in hierarchical databases to support fast malware lookups and comparisons.\\
In this paper we explore the potentials of call graph based malware identification and classification. First call graphs are introduced in more detail as well as graph similarity metrics to compare malware via their call graph representations in Sections 2 and 3. At the basis of call graph comparisons lay graph matching algorithms. Exact graph matchings are expensive to compute, and hence we resort to approximation algorithms (Sections 3, 4). Finally, in Section 5,  the graph similarity metrics are used for automated malware classification via clustering algorithms on a collection of real malware call graphs. A more extensive report on the work is available in \cite{KIN10}.

%



\section{Introduction to Call Graphs}
\label{sec:callgraphintroduction}
A call graph models a binary executable as a directed graph whose vertices, representing the functions the executable is composed of, are interconnected through directed edges which symbolize function calls \cite{callGraph}. A vertex can represent either one of the following two types of functions:
\begin{enumerate}
 \item Local functions, implemented by the program designer.
 \item External functions: system and library calls.
\end{enumerate}
Local functions, the most frequently occurring functions in any program, are written by the programmer of the binary executable.
External functions, such as system and library calls, are stored in a library as part of an operating system. Contrary to local functions, external functions never invoke local functions.\\
Analogous to \cite{hcs09}, call graphs are formally defined as follows:
\begin{mydef}
(Call Graph): A call graph is a directed graph G with vertex set V=V(G), representing the functions, and edge set E=E(G), where E(G) $\subseteq$ V(G)$\times$V(G), in correspondence with the function calls.
\end{mydef}

Call graphs are generated from a binary executable through static analysis of the binary with disassembly tools \cite{malwareCreation}. First, obfuscation layers are removed, thereby unpacking and, if necessary, decrypting the executable. Next, a disassembler like IDA Pro \cite{idapro} is used to identify the functions and assign them symbolic names. Since the function names of user written functions are not preserved during the compilation of the software, random yet unique symbolic names are assigned to them. External functions, however, have common names across executables. In case an external function is imported dynamically, one can obtain its name from the Import Address Table (IAT) \cite{peFileFormat1,peFileFormat2}. When, on the other hand, a library function is statically linked, the library function code is merged by the compiler into the executable. If this is the case, software like IDA Pro's FLIRT \cite{idapro_flirt} has to be used to recognize the standard library functions and to assign them the correct canonical names. Once all functions, i.e.\ the vertices in the call graph, are identified, edges between the vertices are added, corresponding to the function calls extracted from the disassembled executable.

\begin{figure}
	\centering
		\includegraphics[width=\linewidth]{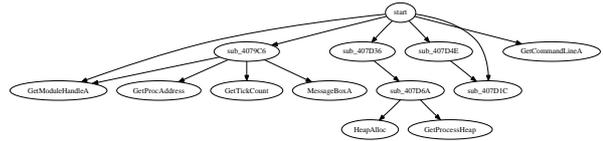}
	\caption[Example of a call graph]{Example of a small malware call graph. Function names starting with 'sub' denote local functions, whereas the remaining functions are external functions.}
	\label{fig:bifrose}
\end{figure}

%
%

\section{Graph Matching}
\label{sec:GraphMatching}


\subsection{Graph matching techniques}
Detecting malware through the use of call graphs requires means to compare call graphs mutually, and ultimately, means to distinguish call graphs representing benign programs from call graphs derived from malware samples. Mutual graph comparison is accomplished with graph matching.
\begin{mydef}
(Graph matching): For two graphs, $G$ and $H$, of equal order, the graph matching problem is concerned with finding a one-to-one mapping (bijection) $\phi: V(G)\rightarrow V(H)$ that optimizes a cost function which measures the quality of the mapping.
\end{mydef}
In general, graph matching involves discovering structural similarities between graphs \cite{approxGraphEdit08} through one of the following techniques:

\begin{enumerate}
	\item Finding graph isomorphisms
	\item Detecting maximum common subgraphs (MCS)
	\item Finding minimum graph edit distances (GED)
\end{enumerate}


An exact graph isomorphism for two graphs, $G$ and $H$, is a bijective function $f(v)$ that maps the vertices $V(G)$ to $V(H)$ such that for all $i,j\in V(G)$, $(i,j)\in E(G)$ if and only if $(f(i),f(j))\in E(H)$ \cite{introductiontoGraph}. Detecting the largest common subgraph for a pair of graphs is closely related to graph isomorphism as it attempts to find the largest induced subgraph of $G$ which is isomorphic to a subgraph in $H$. Consequently, one could interpret an exact graph isomorphism as a special case of MCS, where the common subgraph encompasses all the vertices and edges in both graphs.  Finally, the last technique, GED, calculates the minimum number of edit operations required to transform graph $G$ into graph $H$.
\begin{mydef} (Graph edit distance): The graph edit distance is the minimum number of elementary operations required to transform a graph G into graph H. A cost is defined for each edit operation, where the total cost to transform $G$ into $H$ equals the edit distance.
\end{mydef}
Note that the GED metric depends on the choice of edit operations and the cost involved with each operation. Similar to \cite{comparingStars,approxGraphEdit08,hcs09}, we only consider vertex insertion/deletion, edge insertion/deletion and vertex relabeling as possible edit operations.\\
We can now show that the MCS problem can be transformed into the GED problem. Given is the shortest sequence of edit operations $ep$ which transforms graph $G$ into graph $H$, for a pair of unlabeled, directed graphs $G$ and $H$. Apply all the necessary destructive operations, i.e.\ edge deletion and vertex deletion, on graph $G$ as prescribed by $ep$. The maximum common subgraph of $G$ and $H$ equals the largest connected component of the resulting graph. Without further proof, this reasoning can be extended to labeled graphs. 

For the purpose of identifying, quantifying and expressing similarities between malware samples, both MCS and GED seem feasible techniques. Unfortunately, MCS is proven to be an NP-Complete problem \cite{npCompleteMCS}, from which the NP-hardness of GED optimization follows by the prevous argument (The latter result was first proven in \cite{comparingStars} by a reduction from the subgraph isomorphism problem). Since exact solutions for both MCS and GED are computationally expensive to calculate, a large amount of research has been devoted to fast and accurate approximation algorithms for these problems, mainly in the field of image processing \cite{MCS_ImageCategorization} and for bio-chemical applications \cite{MCS_chem_struc,MCS_protein}. The remainder of this Subsection serves as a brief literature review of different MCS and GED approximation approaches.\\
A two-stage discrete optimization approach for MCS is designed in \cite{MCS_2StageOptim}. In the first stage, a greedy search is performed to find an arbitrary common subgraph, after which the second stage executes a local search for a limited number of iterations to improve upon the graph discovered in stage one. Similarly to \cite{MCS_2StageOptim}, the authors of \cite{MCS_protein} also rely on a two-stage optimization procedure, however contrary to \cite{MCS_2StageOptim}, their algorithm tolerates errors in the MCS matching. A genetic algorithm approach to MCS is given in \cite{MCS_geneticAlg}. Finally, a distributed technique for MCS based on message passing is provided in \cite{MCS_MessagePassing}.\\
A survey of three different approaches to perform GED calculations is conducted by Neuhaus, Riesen, et.\ al.\ in \cite{approxGraphEdit08,approxGraphEdit07, approxGraphEdit06}. They first give an exact GED algorithm using A* search, but this algorithm is only suitable for small graph instances \cite{approxGraphEdit06}. Next, A*-Beamsearch, a variant of A* search which prunes the search tree more rigidly, is tested. As is to be expected, the latter algorithm provides fast but suboptimal results. The last algorithm they survey uses Munkres' bipartite graph matching algorithm as an underlying scheme. Benchmarks show that this approach, compared to the A*-search variations, handles large graphs well, without affecting the accuracy too much. In \cite{GED_BLProgramming}, the GED problem is formulated as a Binary Linear Program, but the authors conclude that their approach is not suitable for large graphs. Nevertheless, they derive algorithms to calculate respectively the lower and upper bounds of the GED in polynomial time, which can be deployed for large graph instances as estimators of the exact GED. Inspired by the work of Justice and Hero in \cite{GED_BLProgramming}, the authors of \cite{comparingStars} developed new polynomial algorithms which find tighter upper and lower bounds for the GED problem.

\subsection{Graph similarity}
In general, malware consists of multiple components, some of which are new and others which are reused from other malware \cite{malwareCreation}. The virus writer will test his creations against several anti-virus products, making modifications along the way until the anti-virus programs do not recognize the virus anymore. Furthermore, at a later stage the virus writer might release new, slightly altered, versions of the same virus \cite{BRMM07,SZOR05}.\\
In this Section, we will describe how to determine the similarity between two malware samples, based on the similarity $\sigma(G,H)$ of their underlying call graphs. As will become evident shortly, the graph edit distance plays an important role in the quantification of graph similarity. After all, the extent to which the malware writer modifies a virus or reuses components should be reflected by the edit distance.

\begin{mydef}
\label{def:graph_similarity}
(Graph similarity): The similarity $\sigma(G,H)$ between two graphs $G$ and $H$ indicates the extent to which graph $G$ resembles graph $H$ and vice versa. The similarity $\sigma(G,H)$ is a real value on the interval [0,1], where 0 indicates that graphs $G$ and $H$ are identical whereas a value 1 implies that there are no similarities. In addition, the following constraints hold: $\sigma(G,H)=\sigma(H,G)$ (symmetry),  $\sigma(G,G)=0$, and $\sigma(G,K_0)=1$ where $K_0$ is the null graph, $G\neq K_0$.
\end{mydef}

Before we can attend to the problem of graph similarity, we first have to revisit the definition of a graph matching as given in the previous Section. To find a bijection which maps the vertex set $V(G)$ to $V(H)$, the graphs $G$ and $H$ have to be of the same order. However, the latter is rarely the case when comparing call graphs. To circumvent this problem, the vertex sets $V(G)$ and $V(H)$ are supplemented with dummy vertices $\epsilon$ such that the resulting sets $V'(G)$, $V'(H)$ are both of size $|V(G)+V(H)|$ \cite{comparingStars,hcs09}. A mapping of a vertex $v$ in graph $G$ to a dummy vertex $\epsilon$ is then interpreted as deleting vertex $v$ from graph $G$, whereas the opposite mapping implies a vertex insertion into graph $H$.
Now, for a given graph matching $\phi$, we can define three cost functions: VertexCost, EdgeCost and RelabelCost.
\begin{description}
\item[VertexCost] The number of deleted/inserted vertices: $|\{v: v\in [V'(G) \cup V'(H)] \wedge[\phi(v)=\epsilon \vee \phi(\epsilon)=v]\}|$.
\item[EdgeCost] The number of unpreserved edges: $|E(G)|+|E(H)|-2\times |\{(i,j):[(i,j)\in E(G) \wedge (\phi(i),\phi(j))\in E(H)]\}|$.
\item[RelabelCost] The number of mismatched functions, i.e.\ the number of external functions in $G$ and $H$ which are mapped against different external functions or local functions.
\end{description}
The sum of these cost functions results in the graph edit distance $\lambda_\phi(G,H)$:  
\begin{equation}
\lambda_\phi(G,H)=VertexCost+EdgeCost+RelabelCost
\label{eq:graphEditDistance}
\end{equation}
Note that, as mentioned before, finding the minimum GED, i.e.\ $\underset{\phi}{min}\;\lambda_\phi(G,H)$, is an NP-hard problem, but can be approximated. The latter is elaborated in the next Subsection.

Finally, the similarity $\sigma(G,H)$ of two graphs is obtained from the graph edit distance $\lambda_\phi(G,H)$:
\begin{equation}
\sigma(G,H)=\frac{\lambda_\phi(G,H)}{|V(G)|+|V(H)|+|E(G)|+|E(H)|}
\label{eq:graphSimilarity}
\end{equation}

\subsection{Graph edit distance approximation}
\label{sec:gedApprox}
Finding a graph matching $\phi$ which minimizes the graph edit distance is proven to be an NP-Complete problem \cite{comparingStars}. Indeed, empirical results show that finding such a matching is only feasible for low order graphs, due to the time complexity \cite{approxGraphEdit06}. In \cite{callGraphSA,KIN10}, the performance of several graph matching algorithms for call graphs is investigated. Based on the findings in \cite{callGraphSA}, the fastest and most accurate results are obtained with an adapted version of Simulated Annealing; a local search algorithm which searches for a vertex mapping that minimizes the GED. This algorithm turns out to be both faster and more accurate than for example the algorithms based on Munkres' bipartite graph matching algorithm as applied in the related works \cite{comparingStars,hcs09}.
Two steps can be distinguished in the Simulated Annealing algorithm for call graph matching. In the first step, the algorithm determines which external functions a pair of call graphs have in common. These functions are mapped one-to-one. Next, the remaining functions are mapped based on the outcome of the Simulated Annealing algorithm, which attempts to map the remaining functions in such a way that the GED for the call graphs under consideration is minimized. For more details, refer to \cite{callGraphSA}.

\section{Clustering}
\label{sec:Clustering}
Writing a malware detection signature for each individual malware sample encountered is a cumbersome and time consuming process. Hence, to combat malware effectively, it is desirable to identify groups of malware with strong structural similarities, allowing one to write generic signatures which capture the commonalities of all malware samples within a group. This Section investigates several approaches to detect malware families, i.e.\ groups of similar malware samples, via clustering algorithms.

\subsection{\emph{k}-medoids clustering}

One of the most commonly used clustering techniques is $k$-means clustering. The formal description of $k$-means clustering is summarized as follows \cite{kmeansPP,patternClassification}:

\begin{mydef}
($k$-means Clustering): Given a data set $\chi$, where each sample $x\in \chi$ is represented by a vector of parameters, $k$-means clustering attempts to group all samples into $k$ clusters. For each cluster $C_i\in C$, a cluster center $\mu_{C_i}$ can be defined, where $\mu_{C_i}$ is the mean vector, taken over all the samples in the cluster. The objective function of $k$-means clustering is to minimize the total squared Euclidean distance $||x-\mu_{C_i}||^2$  between each sample $x\in \chi$, and the cluster center $\mu_{C_i}$ of the cluster the sample has been allocated to:
\[
min \sum_{i=1}^k\sum_{x\in C_i}||x-\mu_{C_i}||^2
\]
\end{mydef}

The above definition assumes that for each cluster, it is possible to calculate a mean vector, the cluster center (also known as centroid), based on all the samples inside a cluster. However, with a cluster containing call graphs, it is not a trivial procedure to define a mean vector. Consequently, instead of defining a mean vector, a call graph inside the cluster is selected as the cluster center. More specifically, the selected call graph has the most commonalities, i.e.\ the highest similarity, with all other samples in the same cluster. This allows us to reformulate the objective function:
\[
min \sum_{i=1}^k\sum_{x\in C_i}{\sigma(x,\mu_{C_i})}
\]
where $\sigma(G,H)$ is the similarity score of graphs $G$ and $H$ as discussed in Section \ref{sec:GraphMatching}. The latter algorithm is more commonly known as a \emph{$k$-medoids} clustering algorithm \cite{kmedoidsPam}, where the cluster centers $\mu_{C_i}$ are referred to as 'medoids'. Since finding an exact solution in accordance with the objective function has been proven to be NP-hard \cite{kMeansNPHard}, an approximation algorithm is used (Algorithm \ref{alg:kmeans}).

\begin{algorithm}
\SetKwComment{Comment}{}{}
\SetCommentSty{emph}
\KwIn{Number of clusters $k$, set of call graphs $\chi$.}
\KwOut{A set of $k$ clusters C}\BlankLine
\ForEach{$C_i\in C$}{
 		Initialize $\mu_{C_i}$ with an unused sample from $\chi$\;
}
\Repeat{The objective function converges}{
Classify the remaining $|\chi|-k$ call graphs. Each sample $x\in \chi$ is put in the cluster which has the most similar cluster medoid\;
\ForEach{$C_i\in C$}{
 		Recompute $\mu_{C_i}$\;
}
}
\Return{ $C={C_0, C_1,...,C_{k-1}}$}
\caption{The $k$-medoids clustering algorithm}\label{alg:kmeans}
\end{algorithm}

In \cite{manning08}, a formal proof on the convergence of $k$-means clustering with respect to its objective function is given. To summarize, the authors of \cite{manning08} prove that the objective function decreases monotonically during each iteration of the $k$-means algorithm. Because there are only a finite number of possible clusterings, the $k$-means clustering algorithm will always obtain a result which corresponds to a (local) minimum of the objective function. Since $k$-medoids clustering is directly derived from $k$-means clustering, the proof also applies for $k$-medoids clustering.\\
To initialize the cluster medoids, we use three different algorithms. The first approach selects the medoids at random from $\chi$. Arthur and Vassilvitskii observed in their work \cite{kmeansPP} that $k$-means clustering, and consequently also $k$-medoids clustering, is a fast, but not necessarily accurate approach. In fact, the clusterings obtained through $k$-means clustering can be arbitrarily bad \cite{kmeansPP}. In their results, the authors of \cite{kmeansPP} conclude that bad results are often obtained due to a poor choice of the initial cluster centroids, and hence they propose a novel way to select the initial centroids, which considerably improves the speed and accuracy of the $k$-means clustering algorithm \cite{kmeansPP}. In summary, the authors describe an iterative approach to select the medoids, one after another, where the choice of a new medoid depends on the earlier selected medoids. For a detailed description of their $k$-means++ algorithm, refer to \cite{kmeansPP}. Finally, the last algorithm to select the initial medoids will be used as a means to assess the quality of the clustering results.  To assist the $k$-medoids clustering algorithm, the initial medoids are selected manually by anti-virus analysts. We will refer to this initialization technique as "Trained initialization".

\subsubsection{Clustering performance analysis}
\label{clusteringPerformanceAnalysis}
\begin{figure}[t]
	\centering
		\includegraphics[width=\linewidth]{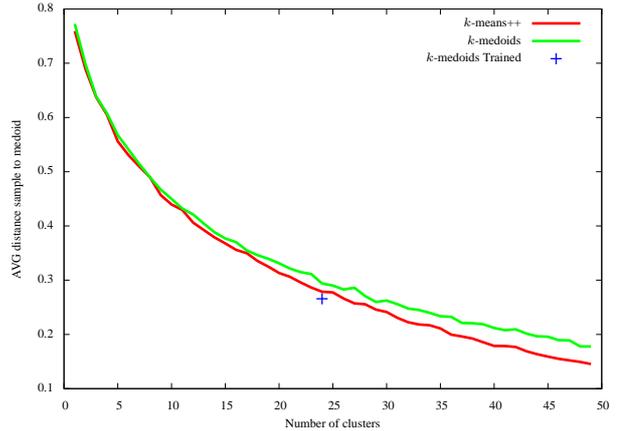}
	\caption[Measuring cluster quality]{Quality of clusters. The average distance $\bar{d}(x_i,\mu_{C_i})$ between a sample $x_i$ in cluster $C_i$ and the cluster's medoid $\mu_{C_i}$ is averaged over 50 executions of the $k$-means algorithm.}
	\label{fig:cclusterQuality_noVertexMatch}
\end{figure}

\begin{figure}[t]
  \centering
\subfloat[][trained $k$-medoids clustering.]{\label{fig:cTrained_noVertexMatching}\includegraphics[width=\linewidth]{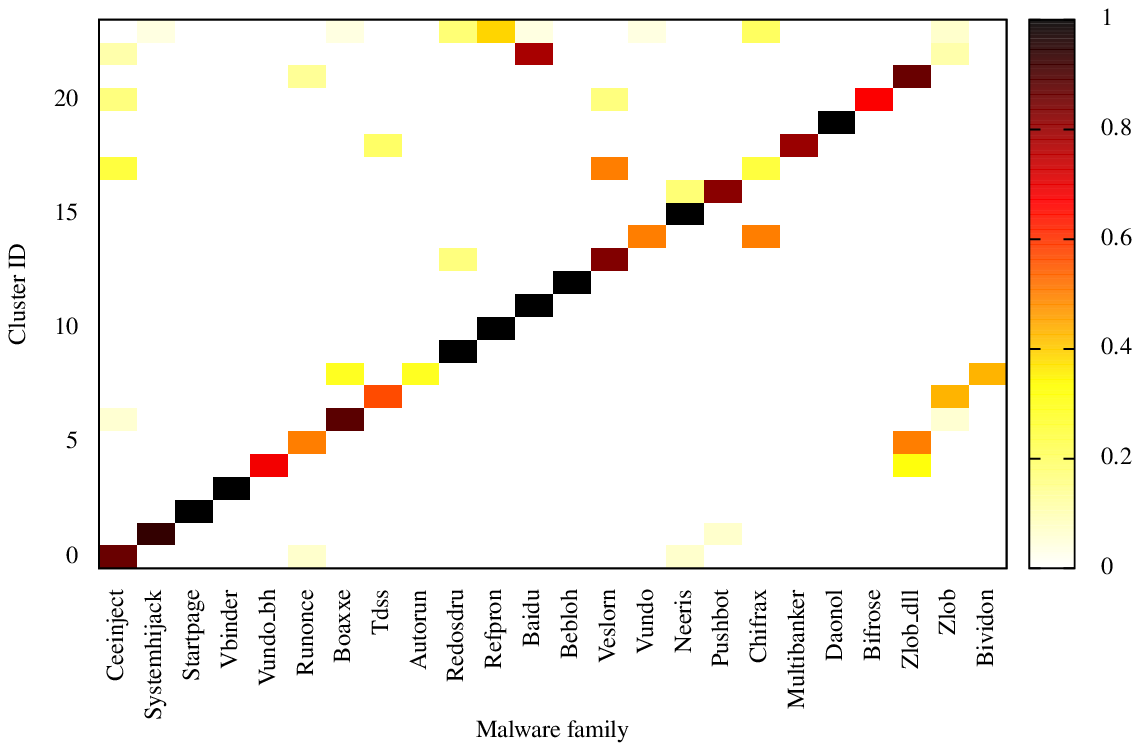}}
\hspace{4mm}
  \subfloat[][$k$-means++ clustering]{\label{fig:ckMeansPP_noVertexMatching}\includegraphics[width=\linewidth]{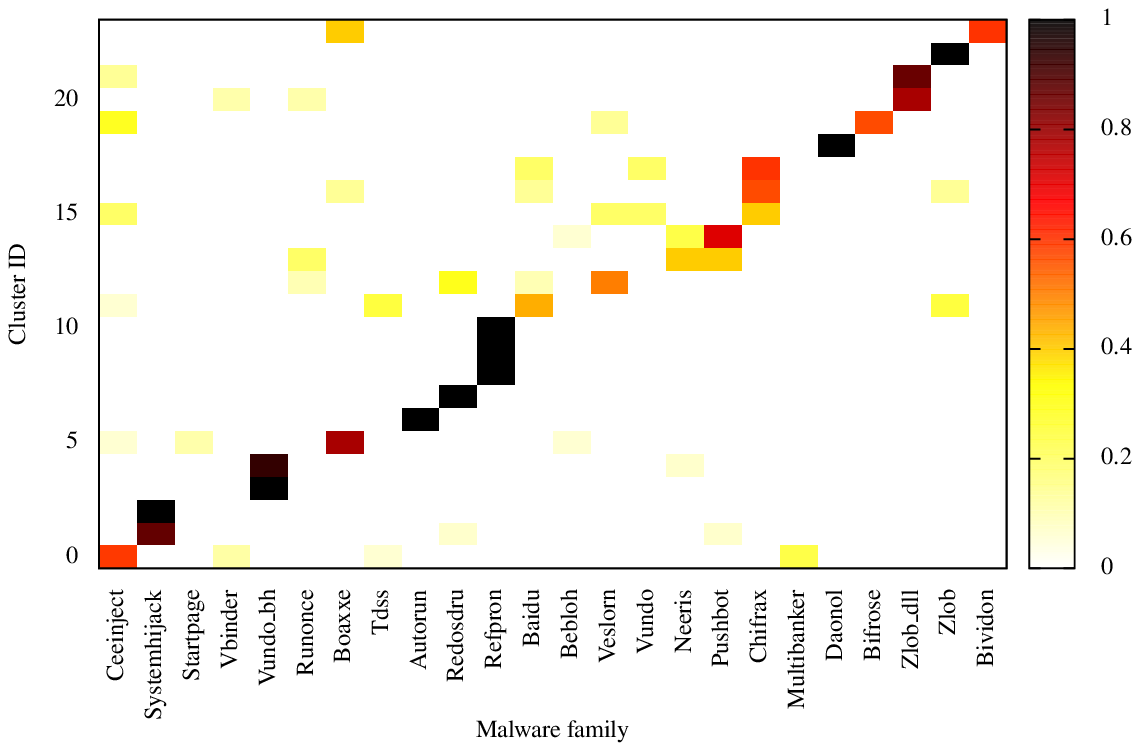}}
  \caption[]{Heat maps depicting non-unique clusterings of 194 samples in 24 clusters. The color of a square depicts the extent to which a certain family is present in a cluster.}
  \label{fig:kMeansClusterHeatmaps}
\end{figure}

In this Subsection, we will test and investigate the performance of $k$-medoids clustering, in combination with the graph similarity scores obtained via the GED algorithm discussed in Section \ref{sec:GraphMatching}. The data set $\chi$ we use consists of 194 malware call graph samples which have been manually classified by analysts from F-Secure Corporation into 24 families. Evaluation of the cluster algorithms is performed by comparing the obtained clusters against these 24 partitions. To get a general impression of the samples, the call graphs in our test set contain on average 234 nodes and 488 edges. The largest sample has 748 vertices and 1875 edges. Family sizes vary from 2 samples to 17 unique call graph samples.\\
Before $k$-medoids clustering can be applied on the data collection, we need to select a suitable value for $k$. Let $k_{optimal}$ be the natural number of clusters present in the data set. Finding $k_{optimal}$ is not a trivial task, and is analysed in depth in the next Subsection. For now, we assume that $k_{optimal}=24$; the number of clusters obtained after manual classification. Note however that $k_{optimal}$ depends on the cluster criteria used to obtain a clustering.
In Figure \ref{fig:cclusterQuality_noVertexMatch}, the average distance $\bar{d}(x_i,\mu_{C_i})$ between a sample $x_i$ in cluster $C_i$ and the medoid of that cluster $\mu_{C_i}$ is plotted against the number of clusters in use. Each time $k$-medoids clustering is repeated, the algorithm could yield a different clustering due to the randomness in the algorithm. Hence, for a given number of clusters $k$, we run $k$-medoids clustering 50 times, and average $\bar{d}(x_i,\mu_{C_i})$. When comparing the different initilization methods of $k$-medoids clustering, based on Figure \ref{fig:cclusterQuality_noVertexMatch}, one can indeed conclude that $k$-means++ yields better results than the randomly initialized $k$-medoids algorithm because $k$-means++ discovers tighter, more coherent clusters. Furthermore, the best results are obtained with Trained clustering where a member from each of the 24 predetermined malware families are chosen as the initial cluster medoids.\\
%
%
Figures \ref{fig:cTrained_noVertexMatching}, \ref{fig:ckMeansPP_noVertexMatching} depict \emph{heat maps} of two possible clusterings of the sample data. Each square in the heat map denotes the presence of samples from a given malware family in a cluster. As an example, cluster 0 in figure \ref{fig:cTrained_noVertexMatching} comprises 86\% Ceeinject samples, 7\% Runonce samples and 7\% Neeris samples. The family names are invented by data security companies and serve only as a means to distinguish families.\\
Figure \ref{fig:cTrained_noVertexMatching} shows the results of $k$-medoids clustering with Trained initialization. The initial medoids are selected by manually choosing a single sample from each of the 24 families identified by F-Secure. The clustering results are very promising: nearly all members from each family end up in the same cluster (Figure \ref{fig:cTrained_noVertexMatching}). Only a few families, such as Baidu and Boaxxe, are scattered over multiple clusters. Figure \ref{fig:ckMeansPP_noVertexMatching} shows the clustering results of $k$-means++ \footnote{A similar figure for randomly initialized $k$-medoids clustering is omitted due to its reduced accuracy with respect to $k$-means++.}. Clearly, the clusterings are not as accurate as with our Trained $k$-medoids algorithm; samples from different families are merged into the same cluster. Nevertheless, in most clusters samples originating from a single family are prominently present. Yet, before one can conclude whether $k$-means++ clustering is a suitable algorithm to perform call graph clustering, one first needs an automated procedure to discover, or at the minimum estimate with reasonable accuracy, $k_{optimal}$. The latter issue is investigated in the next Subsection.

\subsection{Determining the number of clusters}
The $k$-medoids algorithm requires the number of clusters the algorithm should deliver as input. Two quality metrics are used to analyse the natural number of clusters, $k_{optimal}$, in the sample set: Sum of Error and the silhouette coefficient. For a more elaborate discussion, and additional metrics, refer to \cite{KIN10}.

\subsubsection{Sum of (Squared) Error}
The Sum of Error ($SE_p$), measures the total amount of scatter in a cluster. The general formula of $SE_p$ is:
\begin{equation}
SE_p=\sum_{i=1}^{k}\sum_{x\in C_i}(d(x_i,\mu_{C_i}))^p
\label{eq:SS}
\end{equation}
In this equation, $d(x,y)$ is a distance metric which measures the distance between a sample and its corresponding cluster centroid (medoid) as a positive real value. Here we choose $d(x_i,\mu_{C_i})=100\times \sigma(x_i,\mu_{C_i})$.\\
Ideally, when one plots $SE_p$ against an increasing number of clusters, one should observe a quick decreasing $SE_p$ on the interval $[k=1,k_{optimal}]$ and a slowly decreasing value on the interval $[k_{optimal},k=|\chi|]$ \cite{introToDatamining}.

\subsubsection{Silhouette Coefficient}
The average distance between a sample and its cluster medoid measures the cluster \emph{cohesion} \cite{introToDatamining}. The cluster cohesion expresses how similar the objects inside a cluster are. The cluster \emph{separation} on the other hand reflects how distinct the clusters are mutually. An ideal clustering results in well-separated (non-overlapping) clusters with a strong internal cohesion. Therefore, $k_{optimal}$ equals the number of clusters which maximizes both cohesion and separation. The notions of cohesion and separation can be combined into a single function which expresses the quality of a clustering: the \emph{silhouette coefficient} \cite{introToDatamining, silhouette}.\\
For each sample $x_i\in\chi $, let $a(x_i)$ be the average similarity of sample $x_i\in C_k$ in cluster $C_k$ to all other samples in cluster $C_k$:
\[
a(x_i)=
\frac{
	\sum_{
  	x_j\in C_k
  } \sigma(x_i,x_j)
}{
	|C_k|-1
}\quad(x_i\in C_k)
\]
Furthermore, let $b^k(x_i), x_i\notin C_k$ be the average similarity from sample $x_i$ to a cluster $C_k$ which does not accommodate sample $x_i$.
\[
b^k(x_i)=
\frac{
	\sum_{
  	x_j\in C_k
  } \sigma(x_i,x_j)
}{
	|C_k|
}\quad(x_i\notin C_k)
\]

Finally, $b(x_i)$ equals the minimum such $b^k(x_i)$:
\[
b(x_i)=\underset{k}{min}\;b^k(x_i)\quad k\in\{0,1,..,|C|\}
\]

The cluster for which $b^k(x_i)$ is minimal, is the second best alternative cluster to accommodate sample $x_i$. From the discussion of cohesion and separation, it is evident that for each sample $x_i$, it is desirable to have $a(x_i)\ll b(x_i)$ so to obtain a clustering with tight, well-separated clusters.\\
The silhouette coefficient of a sample $x_i$ is defined as:
\begin{equation}
s(x_i)=\frac{b(x_i)-a(x_i)}{max(a(x_i),b(x_i))}
\label{eq:silhouette}
\end{equation}
It is important to note that $s(x_i)$ is only defined when there are 2 or more clusters. Furthermore, $s(x_i)=0$ if sample $x_i$ is the only sample inside its cluster \cite{silhouette}.\\
The silhouette coefficient $s(x_i)$ in Equation \ref{eq:silhouette} always yields a real value on the interval $[-1,1]$. To measure the quality of a cluster, the average silhouette coefficient over the samples of the respective cluster is computed. An indication of the overall clustering quality is obtained by averaging the silhouette coefficient over all the samples in $\chi$. To find $k_{optimal}$, one has to search for a clustering that yields the highest silhouette coefficient.\\
For a single sample $x_i$, $s(x_i)$ reflects how well the sample is classified. Typically, when $s(x_i)$ is close to 1, the sample has been classified well. On the other hand, when $s(x_i)$ is a negative value, then sample $x_i$ has been classified into the wrong cluster. Finally, when $s(x_i)$ is close to 0, i.e.\ $a(x_i)\approx b(x_i)$, it is unclear to which cluster sample $x_i$ should belong: there are at least two clusters which could accommodate sample $x_i$ well.\\

\subsubsection{Experimental results}

\begin{figure}
  \centering \subfloat[][$SS_p$ for various $p$]{\label{fig:malwareClusterSSE}\includegraphics[width=0.42\textwidth]{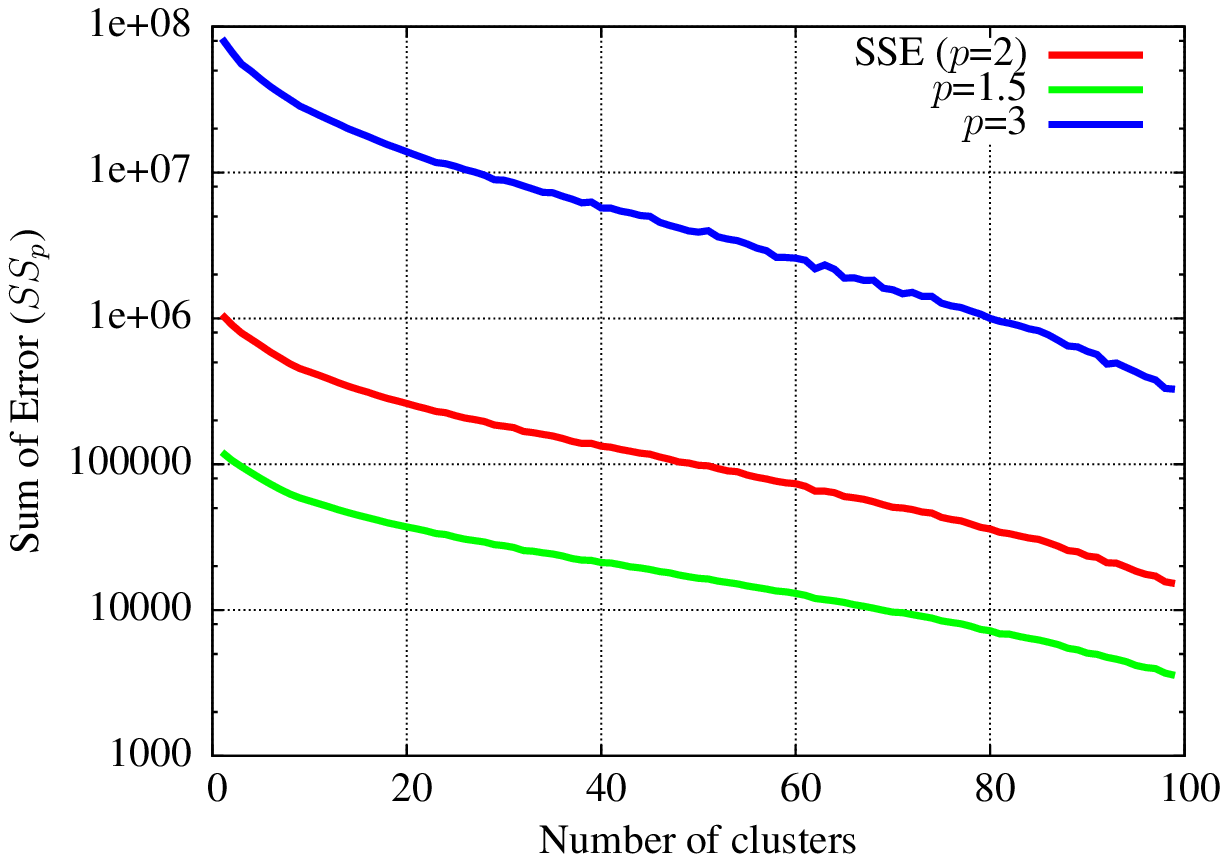}}
  \hspace{4mm}
  \subfloat[][Silhouette Coefficient]{\label{fig:malwareClusterSilhouette}\includegraphics[width=0.42\textwidth]{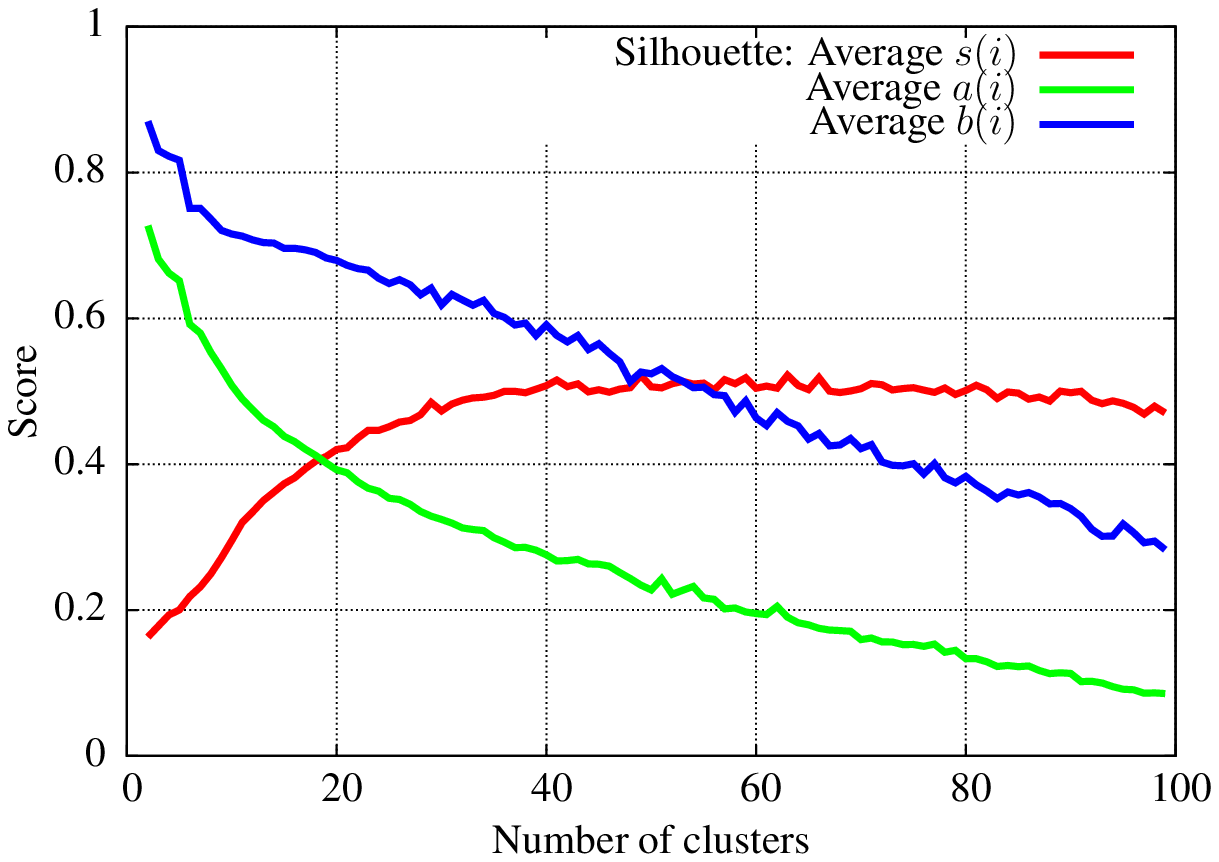}}
  \caption[Finding $k_{optimal}$ in a data set containing malware call graphs.]{Finding $k_{optimal}$ in the set with 194 pre-classified malware samples.}
  \label{fig:malwareCluster}
\end{figure}

The $SE_p$ and silhouette coefficients obtained after clustering the 194 malware samples for various numbers of clusters are depicted in Figure \ref{fig:malwareCluster}. Since the results of the clustering are subject to the randomness in $k$-medoids clustering, each clustering is repeated 10000 times, and the best obtained results are used in Figure \ref{fig:malwareCluster}. Interestingly, the $SE_p$ curves for different values of $p$ in Figure \ref{fig:malwareClusterSSE} do not show an evident value for $k_{optimal}$. Similarly, no clear peak in the silhouette plot (Figure \ref{fig:malwareClusterSilhouette}) is visible either, making it impossible to derive $k_{optimal}$. Consequently, using a $k$-means based algorithm, it is not possible to partition all samples in cohesive, well-separated clusters based on the graph similarity scores, such that the result corresponds with the manual partitioning of the samples by F-Secure. Furthermore, experimental results show that for some samples it is unclear to which cluster they should be assigned too, hence making it difficult to automatically reproduce the 24 clusters as proposed by F-Secure.\\

\subsection{DBSCAN clustering}
In the previous Section, we have concluded that the entire sample collection cannot be partitioned in well-defined clusters, such that each cluster is both tight and well-separated, using a $k$-means based clustering algorithm. Central to the $k$-medoid clustering algorithm stands the selection of medoids. A family inside the data collection is only correctly identified by $k$-medoids if there exists a medoid with a high similarity to all other samples in that family. This, however, is not necessary the case with malware. Instead of assuming that all malware samples in a family are mutually similar to a single parent sample, it is more realistic to assume that malware evolves. In such an evolution, malware samples from one generation are based on the samples from the previous generation. Consequently, samples in generation $n$ likely have a high similarity to samples in generation $n+1$, but samples in generation 0 are possibly quite different from those in generation $n, n\gg 0$. This evolution theory suggests that there are no clusters where the samples are positioned around a single center in a spherical fashion, which makes it much harder for a $k$-means based clustering algorithm to discover clusters.
Although the $k$-medoids algorithms failed to partition all 194 samples in well defined clusters, both Figure \ref{fig:cTrained_noVertexMatching} and Figure \ref{fig:ckMeansPP_noVertexMatching} nevertheless reveal a strong correspondence between the clusters found by the $k$-medoids algorithm and the clusters as predefined by F-Secure. This observation motivates us to investigate partial clustering of the data which discards samples for which it is not clear to which cluster or family they belong. For this purpose, we apply the Density-Based Spatial Clustering of Applications with Noise (DBSCAN) clustering algorithm \cite{introToDatamining, dbscan}. DBSCAN clustering searches for dense areas in the data space, which are separated by areas of low density. Samples in the low density areas are considered noise and are therefore discarded, thereby ensuring that the clusters are well-separated. An advantage of DBSCAN clustering is that the high density areas can have an arbitrary shape; the samples do not necessarily need to be grouped around a single center.\\
To separate areas of low density from high density areas, DBSCAN utilizes two parameters: $MinPts$ and $Rad$. Using these parameters, DBSCAN is able to distinguishes between three types of sample points:
\begin{itemize}
	\item Core points: $P_c=\{x\in \chi, |N_{Rad}(x)|>MinPts\}$, where\\ $N_{Rad}(x)=\{y\in \chi, \sigma(x,y)\leq Rad\}$
	\item Border points: $P_b=\{x\in(\chi\backslash P_c),\exists y\in P_c:\sigma(x,y)\leq Rad\}$
	\item Noise points: $P_n=\chi \backslash (P_c \cup P_b)$
\end{itemize}
An informal description of the DBSCAN clustering algorithm is given in Algorithm \ref{alg:dbscan}.

\begin{algorithm}
\SetKwComment{Comment}{}{}
\SetCommentSty{emph}
\KwIn{Set of call graphs $\chi$, $MinPts$, $Rad$}
\KwOut{Partial clustering of $\chi$}\BlankLine
Classify $\chi$ in Core points, Border points and Noise\;
Discard all samples classified as noise\;
Connect all pairs $(x,y)$ of core points with $\sigma(x,y)\leq Rad$\;
Each connected structure of core points forms a cluster\;
For each border point identify the cluster containing the nearest core point, and add the border point to this cluster\;
\Return{Clustering}
\caption{DBSCAN clustering algorithm}\label{alg:dbscan}
\end{algorithm}

\begin{figure}
  \centering
\subfloat[][Similarity to $k$-nearest neighbor, for different values of $k$, sorted in ascending order. ]{\label{fig:dbscanRad}\includegraphics[width=0.42\textwidth]{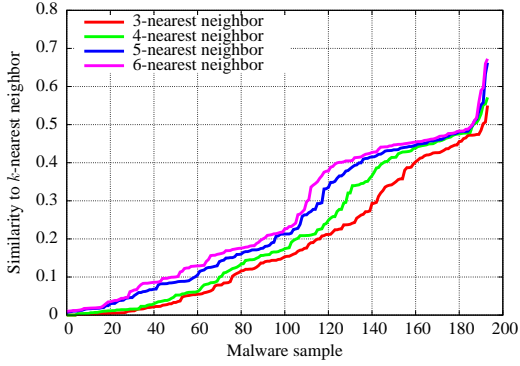}}
\hspace{4mm}
  \subfloat[][Silhouette coefficient for different combinations of $Rad$ and $MinPts$.]{\label{fig:dbscanSilhouette}\includegraphics[width=0.42\textwidth]{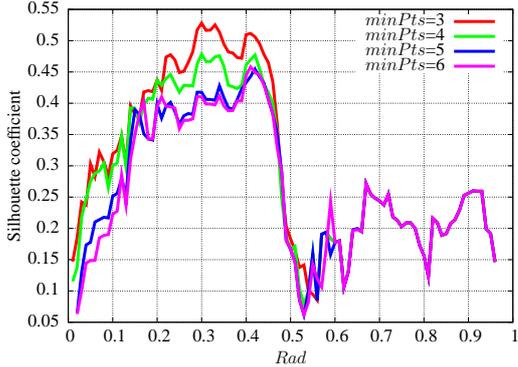}}
  \caption[Discovering necessary parameters for DBSCAN clustering.]{Finding $Rad$ and $MinPts$}
  \label{fig:dbscanParams}
\end{figure}

The question now arises how to select the parameters $MinPts$ and $Rad$. Based on experimental results, the authors of \cite{dbscan} find $MinPts=4$ to be a good value in general. To determine a suitable value for $Rad$, the authors suggest to create a graph where the samples are plotted against the distance (similarity) to their $k$-nearest neighbor in ascending order. Here $k$ equals $MinPts$. The reasoning behind this procedure is as follows: Core or Border points are expected to have a nearly constant similarity to their $k$-nearest neighbor, assuming that $k$ is smaller than the size of the cluster the point resides in, and that the clusters are roughly of equal density. Noise points, on the contrary, are expected to have a relatively larger distance to their $k$-nearest neighbor. The latter change in distance should be reflected in the graph, since the distances are sorted in ascending order.\\
Figure \ref{fig:dbscanRad} shows the similarity of our malware samples to their $k$-nearest neighbors, for various $k$. Arguably, one can observe rapid increases in slope both at $Rad=2.2$ and $Rad=4.8$ for all $k$. A $Rad=4.8$ radius can be considered too large to apply in the DBSCAN algorithm since such a wide radius would merge several natural clusters into a single cluster. Even though $Rad=2.2$ seems a plausible radius, it is not evident from Figure \ref{fig:dbscanRad} which value of $Minpts$ should be selected. To circumvent this issue, DBSCAN clustering has been performed for a large number of $Minpts$ and $Rad$ combinations (Figure \ref{fig:dbscanSilhouette}). For each resulting partitioning, the quality of the clusters has been established with the silhouette coefficient. From Figure \ref{fig:dbscanSilhouette} one can observe that the best clustering is obtained for $Minpts=3$ and $Rad=0.3$. While comparing Figure \ref{fig:dbscanSilhouette} against Figure \ref{fig:dbscanRad}, it is not evident why $Rad=0.3$ is a good choice. We however believe that the silhouette coefficient is the more descriptive metric. 

\begin{figure}[t]
		\includegraphics[width=\linewidth]{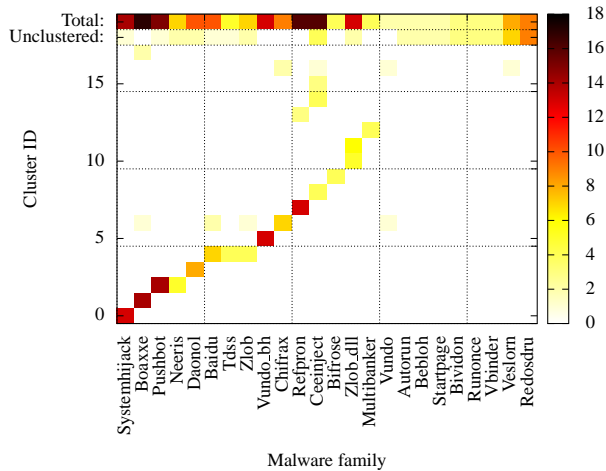}
	\caption[Result of DBSCAN clustering.]{DBSCAN clustering with $Minpts=3$, $Rad=0.3$. The colors depict the frequency of occurrence of a malware sample from a certain family in a cluster.}
	\label{fig:cDbscan_3Minpts_30Rad}
\end{figure}

\begin{figure}[t]
		\includegraphics[width=\linewidth]{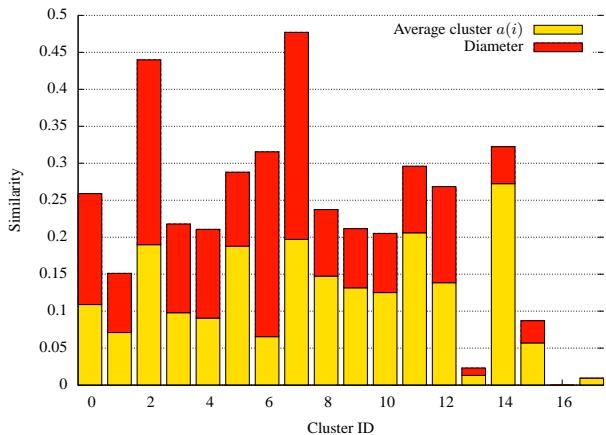}
	\caption[Plot of the diameter and tightness of DBSCAN clustering.]{Plot of the diameter and tightness of the DBSCAN clustering.}
	\label{fig:cdiameter}
\end{figure}

Finally, Figure \ref{fig:cDbscan_3Minpts_30Rad} gives the results of the DBSCAN algorithm for $Minpts=3$ and $Rad=0.3$ in a frequency diagram. Each colored square gives the frequency of samples from a given family present in a cluster. The top two rows of the diagram represent respectively the total size of the family, and the number of samples from a family which were categorized as noise. For example, the Boaxxe family contains 17 samples in total, which were divided over clusters 1 (14 samples), 6 (1 sample), and 17 (2 samples). No samples of the Boaxxe family were classified as noise. The observation that the Boaxxe family is partitioned in multiple clusters is analysed in more detail; closer analysis of this family revealed that there are several samples within the family with a call graph structure which differs significantly from the other samples in the family.\\
The results from the DBSCAN algorithm on the malware samples are very promising. Except 3 clusters, each cluster identifies a family correctly without mixing samples from multiple families. Furthermore, the majority of samples originating from larger families were classified inside a cluster and hence were not considered noise. Families which contain fewer than $Minpts$ samples are mostly classified as noise (e.g.\ Vundo, Blebloh, Startpage, etc), unless they are highly similar to samples from different families (e.g.\ Autorun). Finally, only the larger families Veslorn (8 samples) and Redosdru (9 samples) were fully discarded as noise. Closer inspection of these two families indeed showed that the samples within the families are highly dissimilar from a call graph perspective.\\ 
Finally, Figure \ref{fig:cdiameter} depicts a plot of the diameter and the cluster tightness, for each cluster in Figure \ref{fig:cDbscan_3Minpts_30Rad}. The diameter of a cluster is defined as the similarity of the most dissimilar pair of samples in the cluster, whereas the cluster tightness is the average similarity over all pairs of samples. Most of the clusters are found to be very coherent. Only for clusters 2, 6, and 7, the diameter differs significantly from the average pairwise similarity. For clusters 2 and 6, this is caused by the presence of samples from 2 different families which are still within $Rad$ distance from each other. Cluster 7 is the only exception where samples are fairly different and seem to be modified over multiple generations. Lastly, a special case is cluster 16, where the cluster diameter is 0. The call graphs in this cluster are isomorphic; one cannot distinguish between these samples based on their call graphs, even though they come from different families. Closer inspection of the samples in cluster 16 by F-Secure Corporation revealed that the respective samples are so-called 'droppers'. A dropper is an installer which contains a hidden malicious payload. Upon execution, the dropper installs the payload on the victim's system. The samples in cluster 16 appear to be copies of the same dropper, but each with a different malicious payload. Based on these findings, the call graph extraction has been adapted such that this type of dropper is recognized in the future. Instead of creating the call graph from the possibly harmless installer code, the payload is extracted from the dropper first, after which a call graph is created from the extracted payload.
\section{Conclusion}

In this paper, automated classification of malware into malware families has been studied. First, metrics to express the similarities among malware samples which are represented as call graphs have been introduced, after which the similarity scores are used to cluster structurally similar malware samples. Malware samples which are found to be very similar to known malicious code, are likely mutations of the same initial source. Automated recognition of similarities as well as differences among these samples will ultimately aid and accelerate the process of malware analysis, rendering it no longer necessary to write detection patterns for each individual sample within a family. Instead, anti-virus engines can employ generic signatures targeting the mutual similarities among samples in a malware family.\\
After an introduction of call graphs in Section 2 and a brief description on the extraction of call graphs from malware samples, Section 3 discusses methods to compare call graphs mutually. Graph similarity is expressed via the graph edit distance, which, based on our experiments, seems to be a viable metric. 
To facilitate the discovery of malware families, Section \ref{sec:Clustering} applies several clustering algorithms on a set of malware call graphs. Verification of the classifications is performed against a set of 194 unique malware samples, manually categorized in 24 malware families by the anti-virus company F-Secure Corporation. The clustering algorithms used in the experiments include various versions of the $k$-medoids clustering algorithm, as well as the DBSCAN algorithm. One of the main issues encountered with $k$-medoids clustering is the specification of the desired number of clusters. Metrics to determine the optimal number of clusters did not yield conclusive results, and hence it followed that $k$-means clustering is not effective to discover malware families.\\
Much better results on the other hand are obtained with the density-based clustering algorithm DBSCAN; using DBSCAN we were able to successfully identify malware families.  At the date of writing, automated classification is also attempted on larger data sets consisting of a few thousand samples. However, manual analysis of the results is a time consuming process, and hence the results could not be included in time in this paper. Future goals are to link our malware identification and family recognition software to a live stream of incoming malware samples. Observing the emergence of new malware families, as well as automated implementation of protection schemes against malware families belong to the long term prospects of malware detection through call graphs.

\subsection*{Acknowledgements}
The authors of this paper would like to acknowledge F-Secure Corporation for providing the data required to perform this research. Special thanks go to Pekka Orponen (Head of the ICS Department, Aalto University), Alexey Kirichenko (Research Collaboration Manager F-Secure), Gergely Erdelyi (Research Manager Anti-malware, F-Secure) for their valuable support and many usefull comments.\\
This work was supported by TEKES as part of the Future Internet Programme of TIVIT (Finnish Strategic Centre for Science, Technology and Innovation in the field of ICT).


\balance
\bibliographystyle{IEEEtranS}


\bibliography{sources}


\label{pages-text}
\end{document}